\documentclass[12pt]{article}
\usepackage{epsfig}
\title{Quadratic integrals of motion for the systems of identical particles-quantum case}
\author{Y. Brihaye\\
Faculte des Sciences,\\ Universite de Mons, 7000 Mons, Belgium\\
C. Gonera\thanks{supported by KBN grant 5 P03B06021},  
P. Kosi\'nski$^*$, P. Ma\'slanka$^*$\ 
\\Department of Theoretical Physics II \\
University of {\L}\'od\'z \\
Pomorska 149/153, 90 - 236 {\L}\'od\'z/Poland\\
S. Giller\thanks{supported by KBN grant, No. 5PO3B 06021}\\
Pedagogical University of Czestochowa,\\
 Armii Krajowej 13/15, 42-200 Czestochowa Poland.}
\date{}

\begin{document}
\maketitle
\begin{abstract}
The quantum dynamical systems of identical particles admitting an additional integral quadratic in momenta
 are considered. 
It is found that an appropriate ordering procedure exists which allows to convert the classical
 integrals into
 their quantum 
counterparts. The relation to the separation of variables in Schroedinger equation is discussed.

\end{abstract}

\newpage
\section{Introduction}

It has been shown by Braden \cite{b1} that assuming the permutational symmetry imposes severe restrictions on the form of 
potentials admitting polynomial in momenta integrals of motion. More precisely, he proved that the only system admitting 
third-order integral of motion with position-independent highest-order term is the celebrated CSM model.

 In the previous paper \cite{b2} we studied 
classical systems of identical particles possessing quadratic integrals of motion.
 We obtained the complete classification of such systems. In
the present paper we discuss their quantum versions.

\section{The classical case.}

The main result of ref. \cite{b2} is as follows: assume that\\
(i) the hamiltonian has a natural form 
\begin{eqnarray}
H=\frac{1}{2}\sum_{i=1}^Np_i^2+V(q_1,\;\dots ,\;q_N)\label{w1}
\end{eqnarray}
(ii) $H$\ is translationally invariant
\begin{eqnarray}
\{H,\;P\}=0,\;\;P=\sum_{i=1}^Np_i\label{w2}
\end{eqnarray}
(iii) $H$\ is invariant under the action of the group $S_N$\ of all permutations of canonical variables $q_i,\;p_i,\;i=1,\;\dots ,\;N$.\\
(iv) $H$\ admits at least one integral of motion quadratic in momenta and functionally independent of $H$\ and $P$.\\
Then the general form of potential $V$\ reads
\begin{eqnarray}
V(\underline{q})=\tilde{V}(\underline{q})+U\left(\sum_{i,\;j=1}^N(q_i-q_j)^2\right), \label{w3}
\end{eqnarray}
where $\tilde{V}$\ is a translationally invariant homogeneous function of degree-$2$\ while $U$\ is an arbitrary
 differentiable function of one variable.\\
The form of additional integrals depends on the potential. If $U\equiv 0$\ (or, more generally, a constant)
 any quadratic integral is a linear
combination of the following ones:
\begin{eqnarray}
&&I_1=(\sum_{k=1}^Nq_kp_k)P-2QH\nonumber \\
&&I_2=2(\sum_{k=1}^Nq_k^2)H-(\sum_{k=1}^Nq_kp_k)^2\label{w4}\\
&&I_3=2QP(\sum_{k=1}^Nq_kp_k)-2Q^2H-(\sum_{k=1}^Nq_k^2)P^2\nonumber
\end{eqnarray}
where $Q=\sum_{k=1}^Nq_k$.
The integrals $I_1,\;I_2,\;I_3$\ obey the relation
\begin{eqnarray}
I_1^2+P^2I_2+2HI_3=0\label{w5}
\end{eqnarray}
For nontrivial $U$\ there exists one integral
\begin{eqnarray}
&&I_4=2\left(N(\sum_{k=1}^Nq_k^2)-Q^2\right)\tilde{H}-N(\sum_{k=1}^Nq_kp_k)^2+\nonumber \\
&&+2QP(\sum_{k=1}^Nq_kp_k)-P^2(\sum_{k=1}^Nq_k^2)\label{w6}
\end{eqnarray}
where $\tilde{H}$\ is obtained from $H$\ by replacement $V\rightarrow \tilde{V}$.\\
We have shown that the above integrals are related to the separation of variables
in the H-J equation. More precisely, they appear from the separation of radial variables in Jacobi coordinates. Moreover, 
it has been explained in ref. \cite{b3} how the integrals are related to conformal $sl(2,{\bf R})$\ symmetry.
\section{The quantum case.}

As it is seen from equations (\ref{w4}), (\ref{w6}) the form of integrals cannot be carried over directly to
 the quantum case: 
the ordering problem arises. However, due to natural form of Hamiltonian one can expect that the integrals
 related to separation of variables in H-J equation
will emerge in the quantum case as the separation constants in the Schroedinger equation.\\
Assuming the form (\ref{w3}) of the potential one can find, by trial and error method, the proper ordering for the quantum counterparts of 
$I_1 \div I_4$:
\begin{eqnarray}
&&I_1=\frac{1}{4}\left(\sum_{k=1}^N(q_kp_k+p_kq_k)\right)P+\frac{1}{4}P\left(\sum_{k=1}^N(q_kp_p+p_kq_k)\right)-QH-HQ \nonumber\\
&&I_2=(\sum_{k=1}^Nq_k^2)H+H(\sum_{k=1}^Nq_k^2)-\frac{1}{4}\left(\sum_{k=1}^N(q_kp_k+p_kq_k)\right)^2 \label{w7} \\
&&I_3=\frac{1}{4}\left[\left(\sum_{k=1}^N(q_kp_k+p_kq_k)\right)(QP+PQ)+(QP+PQ)\left(\sum_{k=1}^N(q_kp_k+p_kq_k)\right)\right]+\nonumber \\
&&-Q^2H-HQ^2-\frac{1}{2}(\sum_{k=1}^Nq_k^2)P^2-\frac{1}{2}P^2(\sum_{k=1}^Nq_k^2)\nonumber \\
&&I_4=\left(N(\sum_{k=1}^Nq_k^2)-Q^2\right)\tilde{H}+\tilde{H}\left(N(\sum_{k=1}^Nq_k^2)-Q^2\right)-\frac{1}{4}N
\left(\sum_{k=1}^N(q_kp_k+p_kq_k)\right)^2+\nonumber \\
&&+\frac{1}{4}\left[\left(\sum_{k=1}^N(q_kp_k+p_kq_k)\right)(QP+PQ)+(QP+PQ)\left(\sum_{k=1}^N(q_kp_k+p_kq_k)\right)\right]\nonumber \\
&&-\frac{1}{2}(\sum_{k=1}^Nq_k^2)P^2-\frac{1}{2}P^2(\sum_{k=1}^Nq_k^2).\nonumber
\end{eqnarray}
One can check explicitely that the above expressions are hermitean and commute with the hamiltonian (again, for 
$I_1,\;I_2,\;I_3$\ being the constants of motion, one has to assume $V=\tilde{V}$).
\section{Separation of variables.}

As in the classical case, one can relate the quadratic integrals to the separation of variables; the relevant equation is now 
the Schroedinger equation. We separate the center-of-mass motion and introduce the polar variables in the space of Jacobi
coordinates. Then the integral $I_4$\ arises from the separation of radial variable. \\
Indeed, let us define
\begin{eqnarray}
&&\tilde{q}_i=q_i-\frac{1}{N}Q\nonumber \\
&&\tilde{p}_i=p_i-\frac{1}{N}P\label{w8}\\
&&\rho =\sqrt{\sum_{k=1}^N\tilde{q}_k^2}\nonumber \\
&&p_{\rho}=\frac{1}{2}\sum_{k=1}^N(\frac{\tilde{\rho}_k}{\rho}\tilde{p}_k+\tilde{p}_k\frac{\tilde{\rho}_k}{\rho})\nonumber
\end{eqnarray}
Then the hamiltonian can be written as
\begin{eqnarray}
H=\frac{1}{2N}P^2+\frac{M^2}{2\rho^2}+\frac{1}{2}p^2_{\rho}+\frac{\hbar^2(N-2)(N-4)}{8\rho^2}+\frac{F}{\rho^2}+U(\rho^2)\label{w9}
\end{eqnarray}
with $M^2$\ being the square of angular momentum in the center-of-mass system i.e.
\begin{eqnarray}
M^2=\frac{1}{2}\sum_{i,j=1}^N(\tilde{q}_i\tilde{p}_j-\tilde{q}_j\tilde{p}_i)^2\label{w10}
\end{eqnarray}
We have also used the general form of potentials (\ref{w3}) which inplies that $F$\ is a function of angular variables only. 
From equation (\ref{w9}) we conclude that
\begin{eqnarray}
\rho^2\left(\tilde{H}-\frac{1}{2N}P^2-\frac{1}{2}p^2_{\rho}-\frac{\hbar^2(N-2)(N-4)}{8\rho^2}\right)\label{w11}
\end{eqnarray}
is the operator responsible for separating the radial variable\\
Now, one can check that the expression (\ref{w11}) equals to
\begin{eqnarray}
\frac{1}{2N}I_4-\frac{\hbar^2(N-2)(N-4)}{8}\label{w12}
\end{eqnarray}

\section{$sl(2,{\bf R})$\ symmetry}

The results obtaioned for $U\equiv 0$\ can  be understood from the point of view of $sl(2,{\bf R})$\ symmetry \cite{b3}. Defining
 $X=\sum_{k=1}^Nq_k^2$\ and $Y=\frac{1}{2}\sum_{k=1}^N(q_kp_k+p_kq_k)$\ one easily checks that,
 together with $H$, the above operators span $sl(2,{\bf R})$\ algebra:
\begin{eqnarray}
&&[Y,\;H]=2i\hbar H\nonumber \\
&&[Y,\;X]=-2i\hbar X\label{w13} \\
&&[H,\;Y]=2i\hbar Y\nonumber
\end{eqnarray}
The Casimir operator of this algebra is of course an integral of motion. One readily finds it is equal to $I_2$:
\begin{eqnarray}
I_2=XH+HX-Y^2\label{w14}
\end{eqnarray}
In order to explain the meaning of $I_1$\ we note the following. Any integral of motion $I$\ which is homogeneous function
of natural degree $n$\ provides a highest-weight vector for some representation in the adjoint action of $sl(2,{\bf R})$, 
$[H,\;I]=0,\;[Y,\;I]=i\hbar nI$. Therefore, the next-to-highest vector $J=\frac{1}{i\hbar}[X,\;I]$\ 
evolves linearly in time:
\begin{eqnarray}
\dot{J}=\frac{1}{i\hbar}[J,\;H]=-2nI\label{w15}
\end{eqnarray}
Now, if $I$' is a second integral of degree $n$' it follows from eq.(\ref{w15}) that $n'I'J-nIJ'$\ 
is also an integral of motion. 
Applying this reasoning to the triple $(H,\;Y,\;X)$\ and the doublet $(P,\;Q)$\ one obtains $I_1$.

\end{document}